\documentclass[a4paper,fleqn,usenatbib]{mnras}

\usepackage{newtxtext,newtxmath}

\usepackage[T1]{fontenc}
\usepackage{ae,aecompl}

\usepackage{graphicx}	
\usepackage{amsmath}	
\usepackage{amssymb}	
\usepackage{natbib}

\newcommand{\kn}{\ensuremath{{\rm  kn}}}
\newcommand{\td}{\ensuremath{t_d}}
\newcommand{\tin}{\ensuremath{t_{\rm inj}}}
\newcommand{\hr}{\ensuremath{\mathcal{H}}}

\newcommand{\vrescale}{\ensuremath{v_{-1}}}
\newcommand{\krescale}{\ensuremath{\kappa_1}}
\newcommand{\mrescale}{\ensuremath{M_{\rm ej,-2}}}

\title{GW170817: A Neutron Star Merger in a Mass-Transferring Triple System}

\author[Chang \& Murray]{Philip Chang$^1$\thanks{chang65@uwm.edu} \& Norman Murray$^{2}$
\\
$^1$Department of Physics, University of Wisconsin-Milwaukee, 3135 N Maryland Ave, Milwaukee, WI 53211, USA\\
$^2$Canadian Institute for Theoretical Astrophysics, 60 St. George Street, University of Toronto, ON M5S 3H8, Canada\\
}

\date{Accepted XXX. Received YYY; in original form ZZZ}

\pubyear{2017}

\begin{document}
\label{firstpage}
\pagerange{\pageref{firstpage}--\pageref{lastpage}}
\maketitle


\begin{abstract}
The light curve of GW170817 is surprisingly blue and bright.  Assuming that the event is a binary neutron star merger, we argue that blueness and brightness of the light curve is the result of ejecta that contains an substantial amount of thermal energy.  To achieve this, the ejecta must be reheated at a substantial distance (1 to 2000 solar radii) from the merger to avoid losing the energy to adiabatic cooling.  We show that this reheating can occur if the merger occurs in a hierarchical triple system where the outer star has evolved and filled its Roche lobe. The outer star feeds mass to the inner binary, forming a circumbinary disc, driving the inner binary to merge. Because the outer star fills its Roche lobe, a substantial fraction of the dynamical ejecta collides with the evolved star, reheating the ejecta in the process. We suggest that the process of mass transfer in heirarchical triples tends to form coplanar triple systems such as PSR J0337+1715, and may provide electromagnetic counterparts to binary black hole mergers.
\end{abstract}

\section{Introduction}\label{sec:introduction}

The recent detection of gravitational waves (GWs) from binary black hole mergers \citep{2016PhRvL.116f1102A,2016PhRvL.116x1103A,2017PhRvL.118v1101A,2017arXiv170909660T} was a triumph of GW astrophysics.  However, the lack of detection in other wavebands did not allow the source to be well localized, due to the low spatial resolution of current GW detectors. The lack of an EM counterpart was not too surprising, as binary black holes in vacuum do not produce robust electromagnetic signals.  

Binary neutron star (BNS) mergers or neutron-star black hole mergers can unbind significant amounts of matter and may be bright electromagnetically. Kilonova emission is one possible electromagnetic counterpart that may be detected by optical or near infrared observations \citep[see for instance][]{2016MNRAS.462.4094S,2016ApJ...826L..13A,2016ApJ...824L..24K,2016ApJ...827L..40S} though the area of sky that needs to be search following a GW detection is formidable ($\sim 1000$ deg$^2$)\citep[see for instance][]{2016ApJ...826L..13A}. The search area is reduced by an order of magnitude with three detectors \citep[e.g.][]{2017arXiv170909660T,GW170817}.

Kilonova emission from a BNS merger may arise from two sources (for recent reviews, see \citealt{2016ARNPS..66...23F} and \citealt{2016AdAst2016E...8T}).  First, the dynamical ejecta of the merger would produced a peak luminosity of $10^{40} - 10^{41}\,{\rm ergs\,s}^{-1}$ a few days to a week after the initial merger. Due to the high opacities of the ejecta, the emission is expected to be mainly in the near infrared.  \citep{2013ApJ...774...25K,2013ApJ...775...18B,2013ApJ...775..113T,2017arXiv170809101T,2017arXiv170202990F}. Second, outflows from the merger disc from viscous heating, neutrino heating, and/or nuclear recombinations would produce high $Y_e$ outflows that are mainly Lanthanide-free \citep{2013MNRAS.435..502F,2014MNRAS.441.3444M,2015ApJ...813....2M,2017arXiv170809101T}.  As a result of the substantially lowered opacity, the resulting light curve from these outflows would peak earlier (at about 1 day) and be bluer \citep{2015MNRAS.450.1777K,2014MNRAS.441.3444M,2017arXiv170809101T}.

Up until recently, claims of kilonova detections have been associated with short gamma-ray bursts.  These detections were at best a single detection in the near infrared  \citep{2013Natur.500..547T,2013ApJ...774L..23B,2013ApJ...775L..19J,2015NatCo...6E7323Y,2016NatCo...712898J}.  This situation changed with the GW detection of GW170817 which was detected in both LIGO and the Virgo detectors \citep{GW170817}. The detection triggered a worldwide follow-up campaign.  The relative closeness of GW170817 of $\approx 40\,{\rm Mpc}$ and superior localization of three detectors allowed rapid EM follow-up of galaxies in the 31 square degree field  \citep{SCIENCE}. A number of instruments around the world were able to track the lightcurve of the electromagnetic counterpart, GW170817, from its early detection in the near-UV and optical at 0.5 days to its rapid fade, and the transition of the peak intensity from the optical to the infrared \citep{SCIENCE}.

The early ($0.5$ d) emission of GW170817 was surprisingly bright ($\approx 10^{42}\,{\rm ergs\,s}^{-1}$) and surprisingly blue, but rapidly faded to the infrared by day 2 \citep{SCIENCE}.  The discovery paper modeled the light curve of GW170817 with a ``concordant'' cocoon breakout model that depends on a delayed jet colliding with the dynamical ejecta, thereby generating the later-time kilonova emission \citep{SCIENCE}. The delayed jet accelerates lower opacity ($\kappa \sim 1\,{\rm cm^{2}\,g^{-1}}$) material to velocities that causes it to peakc early and is followed by slower (0.1c) higher opacity ($\kappa \sim 10\,{\rm cm^{2}\,g^{-1}}$) material that forms the bulk of the ejecta. This ``concordant'' model requires an ejecta mass of $\approx 0.05\,{\rm M_{\odot}}$, which is significantly higher than the maximum ejecta mass of $0.01\,{\rm M_{\odot}}$ predicted by state of the art numerical simulations of BNS mergers \citep[see for instance][]{2011ApJ...738L..32G,2013PhRvD..87b4001H,2015PhRvD..91l4041D}.

In this Letter, we propose an alternative to the ``concordant'' model of \citet{SCIENCE}. Here we argue that a combination of substantially preheated ejecta and kilonova emission explains the light curve of GW170817 in \S\ref{sec:light curve}.
In \S \ref{sec:preheating}, we argue that for this preheated emission to be comparable to kilonova emission at about 0.5 days, the preheating must occur at a distance of between $1-2000\,R_{\odot}$  from the initial merger.  We then argue that this preheating may be a natural consequence if the BNS is in a heirarchical triple where the outer star evolves up the RGB or AGB phase and fills its Roche-lobe in \S \ref{sec:triple}. Mass loss from this outer giant creates a circumbinary disc around the BNS, driving it to merge.  After merger, the dynamical ejecta collides with the star, with the resultant shock reheating the adiabatically cooled ejecta material.  We close with a discussion in \S \ref{sec:discussion} where we speculate on the implications of this scenario for PSR J0337+1715 and electromagnetic counterparts of binary black hole mergers.

\section{Analytic Light Curves}\label{sec:light curve}

We model the ejecta as a constant density expanding sphere.  The expansion adiabatically cools the thermal energy, but does not affect the latent heat of radioactive decay.  As a result, the thermal energy from the latter dominates at when the shell reaches large radii, where photons are able to readily escape.  Radiation in an optically thick medium diffuses outward at an effective speed of $v_{\rm diff} = c/\tau$, where $\tau = \kappa \rho R$ is the optical depth, $\kappa$ is the opacity, $\rho$ is the density, and $R = vt$ is the radius of the sphere, and $v$ is the expansion velocity. For a constant density sphere where $\rho = M_{\rm ej}/(4\pi R^3/3)$ and $M_{\rm ej}$ is the ejecta mass, the optical depth scales like
\begin{equation}
 \tau \sim \kappa\rho R\propto R^{-2} = v^{-2} t^{-2}.
\end{equation}
As a result, the bulk of the radiation is trapped at early times when $\tau \gg 1$ and $c/\tau \ll v$.  At late times, when $v_{\rm diff} > v$, the radiation can escape.  For radioactive heating that follows a exponential decay law, e.g., for supernovae, this gives a late time luminosity:
\begin{equation}\label{eq:late}
 L_{\rm late} = \dot{q}M_{\rm ej} = \dot{q}_0 M_{\rm ej}\exp\left(-\frac{t}{\mathcal{T}}\right),
\end{equation}
where $\dot{q}$ is the heating rate per unit mass and $\mathcal{T}$ is the decay timescale.

To derive the early time light curve, we consider $t\ll \mathcal{T}$ so that $\dot{q} \approx \dot{q}_0$ and $\tau(R) \gg 1$.  As noted earlier, the bulk of the radiation is trapped, but radiation near the surface escapes.  So radiation escapes from an optical depth of $c/\tau(\Delta R) = v$, which corresponds a physical depth of
\begin{equation}\label{eq:depth}
 \Delta R = \frac {c}{v \kappa \rho}.
\end{equation}
Radiation down to this depth can escape which gives a total luminosity of
\begin{equation}\label{eq:early}
 L_{\rm early} = \dot{q}\Delta M_{\rm ej}= 4\pi R^2 \rho \Delta R \dot{q} \propto t^2.
\end{equation}
This $t^2$ power law has been use to fit the early rise of SN2011fe \citep{2011Natur.480..344N}.  Here, we have assumed a constant density distribution and an uniform distribution of radioactive material.  Deviations from these assumptions yield different power laws \citep{2013ApJ...769...67P,2014ApJ...784...85P,2016ApJ...826...96P}.

Finally, setting $\Delta R = R$ and $R = v t$ in equation (\ref{eq:depth}) gives the time of peak light:
\begin{equation}
 t_{\rm peak} = \sqrt{\frac{\kappa M_{\rm ej}}{v c}} \approx 5\times 10^5 \krescale^{1/2} \mrescale^{1/2} \vrescale^{-1/2},
\end{equation}
where $\mrescale=M_{\rm ej}/10^{-2}M_{\odot}$, $\vrescale=v/0.1c$ and $\kappa_1=\kappa/1\,{\rm cm^2\,g}^{-1} $.

For kilonova, a major difference is that the exponential decay law of equation (\ref{eq:late}) no longer holds, as there are a large number of different lanthanides with different radioactive decay timescales. \citet{2010MNRAS.406.2650M} found that this gives a heating rate of
\begin{equation}\label{eq:heating rate}
 \dot{q}_{\kn} = \dot{q}_{\kn,0} \td^{-1.3},
\end{equation}
where $\dot{q}_{\kn,0} = 2\times 10^{10}\,{\rm ergs\,s^{-1}\,g^{-1}}$ and $\td=t/1\,{\rm day}$ \citep[see also][]{2014ApJ...789L..39W,2017MNRAS.468...91H,2017arXiv170809101T}. As a result of equation (\ref{eq:heating rate}), equations (\ref{eq:late}) and (\ref{eq:early}) are modified and become
\begin{eqnarray}
 L_{\rm \kn, late} &=& 4\times 10^{41} \mrescale\td^{-1.3}\,{\rm ergs\,s}^{-1}, \\
 L_{\rm \kn, early} &=& 2\times 10^{41} \krescale^{-1} \vrescale \td^{0.7}\,{\rm ergs\,s}^{-1}.
\end{eqnarray}
To smoothly join the early and late-time kilonova light curves, we assume a kilonova light curve of the form
\begin{equation}\label{eq:radioactive}
 L_{\kn} = \left(L_{\rm \kn, early}^{-2} + L_{\rm \kn, late}^{-2}\right)^{-1/2} = L_0\left(\eta_e^{-2} t_d^{-1.4} + {t_d}^{2.6}\right)^{-1/2},
\end{equation}
where $L_0$ and $\eta_e$ parameterize the uncertainties in $\kappa$, $M_{\rm ej}$, $\dot{q}$, and $v$.  Here the peak in the light curve is given by $dL/dt = 0$ or
\begin{equation}
 t_{\rm p} \approx 0.7\eta_e^{-1/2}\,{\rm days}
\end{equation}


We now assume that the thermal energy of the ejecta is substantial. This is in contrast to supernova Ia where the early time shock heated light curve \citep{2010ApJ...708..598P} is rapidly swamped by the rising radioactive nickel luminosity \citep{2014ApJ...784...85P,2016ApJ...826...96P}. The preheated luminosity is
\begin{equation}
 L_{\rm pre} = 4\pi R^2\sigma T_r^4 = L_{\rm pre,0} \td^{-2}
\end{equation}
where the radiation temperature scales like $T_r \propto t^{-1}$ and the ejecta size scales like, $R \propto t$.  Our fitting formula (eq.[\ref{eq:radioactive}]) becomes:
\begin{equation}
    L_{\rm tot} = L_{\kn} + L_{\rm pre} = L_0\left[\left(\eta_e^{-2} \td^{-1.4} + \td^{2.6}\right)^{-1/2}+\eta_s \td^{-2}\right],
\end{equation}
where $\eta_s = L_{\rm pre,0}/L_0$ encapsulates our ignorance of the properties of the preheated material.

\begin{figure}
 \includegraphics[width=0.45\textwidth]{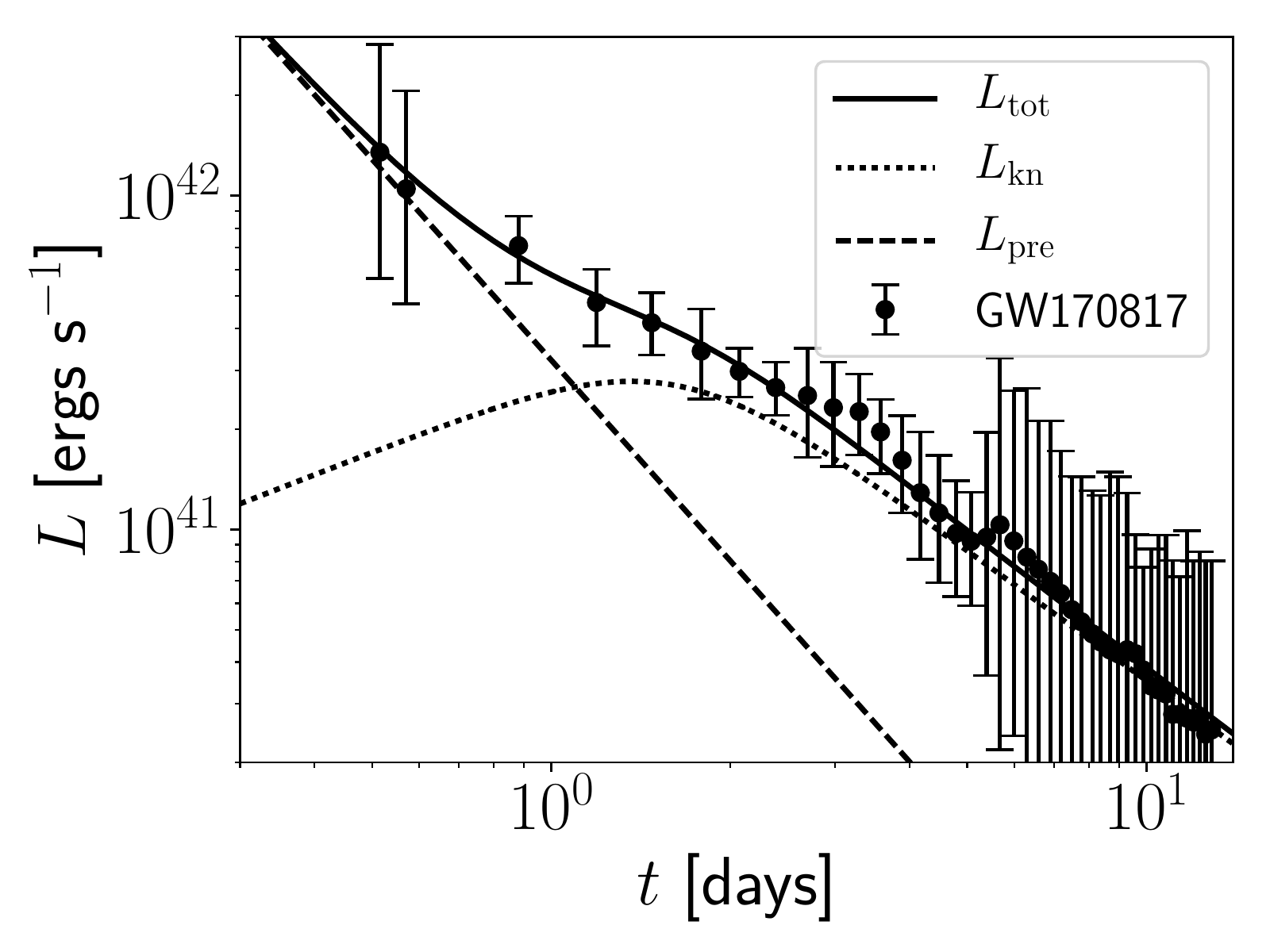}
 \caption{The total luminosity, $L_{\rm tot}$ (solid line), kilonova luminosity, $L_{\kn}$ (dotted line), and preheating luminosity, $L_{\rm pre}$ (dashed line), as a function of time in days for $L_0 = 7\times 10^{41}\,{\rm ergs\,s}^{-1}$, $\eta_e = 0.39$, and $\eta_s = 0.46$.  These parameters were chosen to fit the broadband data for GW170817 from \citet{SCIENCE}, which is shown by the solid points and error bars. 
 \label{fig:example}}
\end{figure}

Figure \ref{fig:example} shows $L_{\rm tot}$ (solid line),  $L_{\kn}$ (dotted line), and  $L_{\rm pre}$ (dashed line) as a function of time for $L_0=7\times 10^{41}\,{\rm ergs\,s}^{-1}$, $\eta_e = 0.39$, and $\eta_s = 0.46$.  These parameters were selected to best fit the data for GW170817 (filled circles with error bars), extracted from Figure 2 of \citet{SCIENCE}.  
Because $\eta_e$ and  $\eta_s$ are both of order unity, neither $L_{\rm pre}$ nor $L_{\kn}$ dominates the total luminosity.  In fact, $L_{\rm pre}$ dominates the emission at very early times, while $L_{kn}$ does so at late times, with the total luminosity approximating a power law.

\section{Constraints on Preheating}\label{sec:preheating}

We now speculate on the source of the preheating in BNS merger ejecta.
We begin with an estimate of the energy budget.  Given the characteristic luminosity of GW170817 at 0.5 days of $L=10^{42}\,{\rm ergs\,s}^{-1}$ \citep{SCIENCE}, the total thermal energy is
\begin{equation}
  E_{\rm Th} \approx 4.3\times 10^{46}L_{42}\left(\frac{\td}{0.5}\right)\,{\rm ergs},
\end{equation}
where $L_{42} = L/10^{42}\,{\rm ergs\,s}^{-1}$. The kinetic energy associated with kilonova ejecta is
\begin{equation}
   E_{\rm kin} = \frac 1 2 m v^2 \approx 10^{50}\left(\frac {M_{\rm ej}}{10^{-2} M_{\odot}}\right)\vrescale^2\,{\rm ergs},
\end{equation}
 with velocities $v\sim 0.1$ c.

At 1 day, the radius of the ejecta is $R = vt \approx 2.6\times 10^{14} \vrescale\td$.  If the kinetic energy of the ejecta sets the energy budget of the trapped radiation, then a lower limit on the radius at which preheating occurs is:
\begin{equation}
\tin > \frac{E_{\rm Th}}{E_{\rm kin}} t \approx 22\,L_{42}\left(\frac {M_{\rm ej}}{10^{-2} M_{\odot}}\right)^{-1}\vrescale^{-2}\left(\frac{\td}{0.5}\right)\,{\rm s},
\end{equation}
which works out to be a distance of
\begin{equation}
  R_{\rm inj} > R^-_{\rm inj} = v \tin = 7\times 10^{10}L_{42}\left(\frac {M_{\rm ej}}{10^{-2} M_{\odot}}\right)^{-1}\vrescale^{-1}\,{\rm cm}.
\end{equation}
The fact that any shock that produces the radiation visible at 0.5 days must occur on a scale greater than 1 solar radii eliminates any injection during the formation of the initial ejecta, i.e., from shocks during the merger process.

The observation of preheated material at 0.5 days sets an upper limit on the preheating radius of:
\begin{equation}
  R_{\rm inj} < R^+_{\rm inj} = v t\approx 1.3\times 10^{14}\vrescale\left(\frac{\td}{0.5}\right)\,{\rm cm},
\end{equation}
which is approximately $2000$ solar radii.

There are a number of ways to reheat the ejecta at these large radii.  First, the ejecta can be reheated by a substantially delayed (in local dynamical times) jet \citep{2017arXiv170510797G,SCIENCE}.  Second, the ejecta can collide with material around the merging neutron star.  The mass of material required would be substantial, i.e., similar to the mass of ejecta itself, which precludes any sort of stellar outflow from a progenitor.  However, we discuss another possibility, that the ejecta collides with a Roche filling third body which both provides a target, and a mechanism that produces the merging BNS.

\section{Giants in a Triple System}\label{sec:triple}

Red Giant Branch (RGB) and Asymptotic Giant Branch (AGB) stars have scales up to $1000$ solar radii.  If a BNS merger occurs in a triple system, then the ejecta from the merger would impact the third star, shock, and reinject thermal energy into the ejecta.  
The kinetic energy and momenta carried by the ejecta, as estimated above, are similar to those carried by supernova ejecta, although the mass of the ejecta in the latter case is much larger than in a neutron star merger. Three dimensional hydrodynamic simulations of the interaction between high velocity ejecta and stars, e.g., \citet{2012A&A...548A...2L}, show that while the companion or target star, with mass $M_c$ and radius $R_c$, will survive, it will lose a significant amount of mass. The simulations agree with simple scaling models for the amount of mass loss \citep{1970Natur.225..247C,1975ApJ...200..145W}. The energetics and timing of the optical emission seen in GW170817 require that some of the stripped and ablated material is accelerated to velocities comparable to that of the ejecta ($v_{\rm ab}\sim0.1 c$). The ratio of the ablated mass that reaches $v_{\rm ab}$ to initial mass of the companion star is $F_{\rm fast} = ({\Sigma_{\rm ej}}/{\Sigma(x_{\rm ab})}) ((v_{\rm ej}/v_{\rm ab})-1)$, where $\Sigma_{\rm ej}$ is the surface density of the ejecta when it encounters the companion star, and $\Sigma(x_{\rm ab})$ is the surface density of the material ablated from the companion that reaches $v_{\rm ab}$.

The ejecta mass $m_{\rm ej}\approx 0.01M_\odot$, so only the outer layers of the companion star, comprising a mass comparable to $m_{\rm ej}$, can be accelerated to $v_{\rm ab}$. The corresponding initial dimensionless cylindrical radius (or impact parameter) $x_{\rm ab}\equiv b/R_c$ of the high velocity ablated material is of order unity. Then the fraction of ablated mass is
\begin{equation}
F_{\rm fast}\approx \frac{1}{4}
\frac{M_{\rm ej}} {M_{\rm c}}
\frac{R^2} {a_{\rm triple}^2}x_{\rm ab}^2,
\end{equation}
where $a_{\rm triple}$ is the separation between the center of mass of the binary and the companion star. We will argue that $a_{\rm triple}\approx 2R_c$, so that $F_{\rm fast}\approx 10^{-3}$.

How might a configuration of a merging BNS with a RGB or AGB star companion in such a tight orbit arise?  The binary fraction of massive stars is large, i.e., of order 70\% \citep[see for instance][]{2013ARA&A..51..269D,2012Sci...337..444S}.  Moreover, a substantial fraction of these stars are in triple or higher multiplicity systems.  Hence it is likely that many BNS may have initially formed in a triple system, and there are a number of known neutron stars that are currently in triple systems, e.g., \citealt{1999ApJ...523..763T,2014Natur.505..520R}.

Hierarchical triple systems may be subject to secular effects such as the Kozai oscillation \citep{1962AJ.....67..591K,1962P&SS....9..719L}. %
%
However, the fraction of triples with the required high inclinations is small, so instead, let us consider the evolution of a third star around a BNS.  As the star goes up the RGB or the AGB, it will fill its Roche lobe and begin to transfer mass to the BNS.  For stars between 1-3 solar masses, this mass transfer is stable, as the donor is lower mass than the accreting BNS.  For stars larger than $\sim 3M_\odot$, mass transfer is unstable and the system enters a common envelope phase, shrinking the orbit of the BNS and the third star.

For stable mass transfer, a circumbinary disc would form around the BNS.  The rate at which angular momentum is transported through the disc is then
\begin{equation}
\dot{L} = \dot{M}\sqrt{GMr_d},
\end{equation}
where $\dot{M}$ is the mass accretion rate onto the disc, $ M$ is the total mass of the BNS, $r_d\lesssim R_{\rm RGB}$ is the size of the disc, and $R_{\rm RGB}$ is the size of the RGB or AGB star, which is similar to the size of the disc assuming that the star fills its Roche lobe.

When the gas in the disc gets down to $r\approx 2 a$, where $a\ll r_d$ is the semimajor axis of the BNS, the disc is truncated \citep{2002ApJ...567L...9A,2008ApJ...672...83M,2010MNRAS.407.2007C} and the gas piles up at that radius.\footnote{\citet{2013MNRAS.436.2997D} \citep[see also][]{2015MNRAS.447L..80F} showed that accretion is not completely arrested, but is reduced by a factor of a few compared to a system without a central binary.}  As a result the orbit of the BNS shrinks on a timescale given by
\begin{equation}
t_{\rm ins} = \frac{M}{\dot{M}}\sqrt{\frac a {r_d}} = 3\times 10^7\left(\frac{M}{3\,M_{\odot}}\right)\dot{M}_{-8}\left(\frac{a/r_d}{10^{-2}}\right)^{1/2}\,{\rm yrs},
\end{equation}
where $\dot{M}_{-8} = \dot{M}/10^{-8} M_{\odot}\,{\rm yr}^{-1}$. Another way to interpret this is that the total mass accreted, $M_{\rm acc}$, to drive the BNS to merge is
\begin{equation}
\frac{M_{\rm acc}}{M} = \sqrt{\frac{a}{r_d}} = 0.1 \left(\frac{a/r_d}{10^{-2}}\right)^{1/2}.
\end{equation}
In any case, the timescale to merger or amount of mass loss required to drive a BNS to merge is shorter than the lifetime of the RGB or AGB star or the total mass of the RGB or AGB envelope.

Typical RGB or AGB stars reach scales up to a few hundred solar radii, but spend a majority of their time at smaller radii.  This would give a semimajor axis of the outer orbit $a_{\rm triple} = 2 R_c \sim 100-1000\,R_{\odot}$. As it is Roche-filling, only 1/16 of the ejecta hits the star.  We can also expect a dilution of the thermal energy of factor of about 10 when it is observed at 0.5 days at a radii of about 2000 $R_{\odot}$.  Thus, we expect an upper limit for the preheated emission of $10^{-2}$ of the kinetic energy.
Finally, while it may seem surprising that a heirarchical triple remains bound given the typically high natal kicks of young pulsars ($\sim 500\,{\rm km\,s}^{-1}$), double NSs typically favor small kicks ($\lesssim 50-100\,{\rm km\,s}^{-1}$ \citealt{2017ApJ...846..170T}). 

\section{Discussion and Conclusions}\label{sec:discussion}

In this Letter, we argue that appropriately preheated ejecta produces a rapidly falling light curve that is initially blue and bright.  Combined with the radioactively heated material from a kilonova, we model the entire light curve of GW170817.  We estimate the scale where this preheating occurs to be between $1-2000\,R_{\odot}$, which excludes any process during the BNS merger.  We argue that a Roche filling RGB or AGB star at this scale is a plausible scenario.  We also argue that mass loss from the RGB or AGB star would form a circumbinary disc that drives the system to merge in the first place.  Therefore, Roche-lobe filling giants in a heirarchical triple may be an important channel for producing merging BNSs and modifying their lightcurves.

Such a triple system is reminiscent of PSR J0337+1715, a millisecond pulsar in a heirarchical triple system \citep{2014Natur.505..520R}. PSR J0337+1715 is a coplanar system, which we suggest may be the result of the outer star having filled its Roche lobe, producing a circumbinary disc that drove the inner binary to align with the outer orbital plane.  In fact, \citet{2014ApJ...781L..13T} argue at one point in the history of the system, the outer star overflowed its Roche lobe, transferring mass  to the inner binary. However, \citet{2014ApJ...781L..13T} do not comment on the possibility that this mass transfer could explain the coplanarity of the inner and outer orbits.

Circumbinary discs have been proposed in several contexts to produce binaries that are interesting from the viewpoint of GWs.  For instance, they have been used to solve the ``final parsec'' problem for supermassive binary black hole (SMBBH) mergers  \citep{2002ApJ...567L...9A,2008ApJ...672...83M,2010MNRAS.407.2007C}.  More recently, they have been proposed to solve the ``final AU'' problem for merging stellar mass black holes \citep{2017MNRAS.464..946S}.   \citet{2017MNRAS.464..946S} proposed that stellar mass binary black holes in AGN discs would develop circumbinary discs that would drive the pair to merge.  Our proposed scenario would be an example of such a circumbinary disc that forms in the field, not in a gas rich environment like an AGN disc, and that produces an observable that is associated with the merger.

Finally, we comment on the circumbinary disk scenario for binary black hole mergers.  The presence of a circumbinary disc would allow for the possibility of an electromagnetic counterpart for which there are two possibilities.  First, the merger of a binary black hole results in both mass loss, via GWs, and a merger kick. Either, or the combination, results in disc shocking as the circumbinary disc suddenly finds itself in non-circular crossing orbits, as in the case of SMBBH mergers \citep{2005ApJ...622L..93M,2008ApJ...676L...5L,2008ApJ...684..835S,2009ApJ...700..859O}.  Second, the circumbinary disc would begin to accrete onto the newly formed merged black hole after a viscous time \citep{2005ApJ...622L..93M}.

To estimate the luminosity in each case, we first calculate the radius at which the binary decouples from the viscous circumbinary disc due to the emission of GWs.  To do so, we set $ t_{\rm visc} = t_{\rm merge}$, where $t_{\rm visc} = \alpha^{-1} \sqrt{r^3/GM} (r/h)^2 $, is the disc viscous timescale at $r$ and $t_{\rm merge} \sim a^4/c r_g^3$ is the merger timescale due to GW losses, where $a$ is the semimajor axis of the inner binary.  Setting $r=2a$, we find the semimajor axis at decoupling,
\begin{eqnarray}
  a_{\rm dec} = 2^{13/5} \frac{r_g}{\alpha^{0.4}}\left(\frac h r\right)^{-0.8}
    \approx 3.6\times 10^{9} M_{50}\alpha_{-2}^{-0.4}\hr_{-1}^{-0.8}\,{\rm cm},
\end{eqnarray}
where $\hr_{-1} = (h/r)/0.1$, $\alpha_{-2} = \alpha/10^{-2}$, and $M_{50} = M/50M_{\odot}$.
At this radius, the order of magnitude luminosity is
\begin{eqnarray}
  L &=& \frac{\Delta M}{M} \frac{GM}{r}\tau_{\rm th}^{-1}\pi\Sigma r^2\nonumber\\
  &\approx&  10^{39} M_{50} \alpha_{-2}^{1.2}\hr_{-1}^{0.4}\Sigma_5\left(\frac{\Delta M/M}{0.05}\right) \,{\rm ergs\,s}^{-1}
\end{eqnarray}
where $\tau_{\rm th} = \alpha^{-1}\sqrt{r^3/GM}$ is the thermal time of the disc and $\Sigma = 10^5\Sigma_5\,{\rm g\,cm}^{-2} $ is the surface density of the disc. The effective temperature  would be a few $\times 10^5$ K.  

For the second possibility, the luminosity would be approximately Eddington, $L \approx 7\times 10^{39} M_{50}\,{\rm ergs\,s^{-1}}$,
with a delay of the viscous time at decoupling, $t_{\rm visc,dec} = t_{\rm visc}(r=2a_{\rm dec}) \approx 21 M_{50}\alpha_{-2}^{-1.6}\hr_{-1}^{-3.2}\,{\rm hrs}$,
which is about a day. The effective temperature of the emission would be akin to a X-ray binary accreting at Eddington,  $T_{\rm eff}\sim 3\times10^6-10^7$ K.

\section*{Acknowledgements}
We thank D. Kaplan for useful conversations.  PC is supported by the NASA ATP program through NASA grant NNX13AH43G and NSF grant AST-1255469. NM was supported in part by the Natural Sciences
and Engineering Council of Canada.

\bibliographystyle{mnras}
\bibliography{kilonova}

\bsp	
\label{lastpage}

\end{document}